\documentclass[%
 reprint,
 amsmath,amssymb,
 aps,
 prl,
]{revtex4-1}

\usepackage{siunitx}
\usepackage{times}
\usepackage{soul}

\usepackage[utf8]{inputenc}

\usepackage{graphicx}
\usepackage{dcolumn}
\usepackage{bm}

\usepackage[caption=false]{subfig}

\usepackage[normalem]{ulem}
\usepackage{color}
\newcommand{\red}[1]{{\color{black}#1}}

\def\eps{\varepsilon}
\def\epsinf{\varepsilon_{\infty}}
\def\w{\omega}
\def\W{\Omega}
\def\wp{\omega_p}
\def\wc{\omega_c}
\def\ws{\omega_s}
\def\g{\gamma}
\def\epst{\Bar{\Bar\varepsilon}}
\def\c{\chi_{\pm}}
\def\c0{\chi_0(\w)}

 \def\Xint#1{\mathchoice
 {\XXint\displaystyle\textstyle{#1}}%
 {\XXint\textstyle\scriptstyle{#1}}%
 {\XXint\scriptstyle\scriptscriptstyle{#1}}%
 {\XXint\scriptscriptstyle\scriptscriptstyle{#1}}%
 \!\int}
\def\XXint#1#2#3{{\setbox0=\hbox{$#1{#2#3}{\int}$}
 \vcenter{\hbox{$#2#3$}}\kern-.5\wd0}}

\def\dashint{\Xint-}

\begin{document}


\title{Broadband and giant nonreciprocity at the subwavelength scale in magnetoplasmonic materials }

\author{Mohamed Ismail Abdelrahman}
 \email{mia37@cornell.edu}

\author{Francesco Monticone}
\email{francesco.monticone@cornell.edu}
\affiliation{Cornell University, School of Electrical and Computer Engineering, Ithaca, New York 14853, USA}

\begin{abstract}
We unveil a previously overlooked wave propagation regime in magnetized plasmonic (gyrotropic) materials with comparable plasma and cyclotron frequencies, which enables a giant and broadband (nondispersive) nonreciprocal response. 
We show that this effect \red{ is due to a natural form of dispersion compensation that ultimately originates from the subtle implications of the principle of causality} for gyrotropic plasmonic media, which allows the existence of a low-loss frequency window with anomalous nonmonotonic dispersion for the extraordinary mode. This is in stark contrast with conventional nongyrotropic passive materials, for which the frequency derivative of the permittivity dispersion function is always positive in low-loss regions.
%
%
These findings  pave the way for superior nonreciprocal components in terms of bandwidth of operation and compactness, with orders-of-magnitude reductions in size. As a relevant example, we consider indium antimonide (InSb) to theoretically demonstrate a deeply subwavelength, broadband, THz isolator operating under moderate magnetic bias and
at room temperature. 
\end{abstract}

\maketitle


\section{I. Introduction}

Breaking Lorentz reciprocity in electromagnetic systems enables a plethora of  complex devices with rich functionalities to control wave propagation in anomalous ways, such as circulators, nonreciprocal phase shifters, asymmetric polarization converters, and isolators. Nonreciprocal effects can be accomplished using various techniques, for example, by exploiting the precession motion of spinning electrons or the cyclotron motion of free electrons  in ferrites and plasma media, respectively, under a static magnetic field \cite{sommerfeld1954lectures,lax1962microwave,pozar2009microwave}. Alternatively, in recent years, magnetic-free techniques have been extensively investigated by leveraging other forms of bias that break time-reversal symmetry, as in artificially engineered metamaterials biased by linear/angular momentum \cite{yu2009complete,sounas2013giant}  or, more generally, in spatio-temporally modulated media \cite{wang2012non,caloz2019spacetime,sounas2017non}. Nonlinear effects have also been exploited to achieve nonreciprocity \cite{gallo2001all,fan2012all}, yet of a weaker form, and limited to applications in which the device is not simultaneously excited from both sides (dynamic nonreciprocity constraints \cite{shi2015limitations}).  
We refer the interested reader to Ref. \cite{caloz2018electromagnetic} for a comprehensive review of nonreciprocal effects and devices.

Within this context, achieving giant nonreciprocity in \emph{broadband} and \emph{compact} devices is an active area of research with several applications at microwaves, terahertz  (THz), and optical frequencies \red{\cite{pozar2009microwave,zvezdin1997modern,arikawa2012giant,mu2019tunable,zhang2019monolithic}}. Particularly appealing are emerging applications in the so-called ``THz gap,'' which extends from the high-end of the millimeter-wave band to the low-end of the far-infrared light band \cite{pawar2013terahertz}, such as ultra-broadband communications, advanced THz imaging and spectroscopy \red{using ultrashort pulses}, and THz microscopy  for material and device characterization \cite{chan2007imaging,piesiewicz2007short,pawar2013terahertz,mittleman2013sensing}.
\red{For many of these applications across the electromagnetic spectrum, nonreciprocal elements are becoming increasingly important to realize required functionalities for propagation/polarization control that are unattainable using only reciprocal components, such as unidirectional polarization conversion and phase shifting, and unidirectional signal transmission and routing. Notably, unidirectional transmission devices, namely, isolators, are  crucial to protect electromagnetic sources and amplifiers from detrimental back reflections. Since many electromagnetic and photonic systems deal with broadband signals of ever-increasing bandwidth, as in the case of communications systems or for applications that rely on ultrashort pulses, it becomes particularly important to realize nonreciprocal devices that operate over the broadest possible bandwidth.} 

%

Attempts to realize broadband isolators rely on many different mechanisms, each with its own limitations. Examples include wavelength-insensitive isolators based on dispersion compensation using additional optical elements \cite{schulz1989wavelength,dimova2013efficient}, which are, however not suitable for integration in monolithic systems due to their bulkiness; broadband nonlinear effects using multiple resonators \cite{sounas2018broadband}, which suffer from the fundamental limitations and constraints mentioned above; and time-modulated microwave/millimeter-wave circuits based on balanced delay lines with switches \cite{dinc2017synchronized}, which are, however, difficult to scale beyond the millimeter-wave range. Historically, the most common approach to realize broadband nonreciprocal devices at {\red{microwave}} frequencies has been to use ferrimagnetic materials (ferrites and magnetic garnets) under static magnetic bias, which naturally exhibit a dispersion-less broadband region of Faraday rotation \cite{lax1962microwave,shalaby2013magnetic}. The operation of ferrite-based {\red{microwave}} isolators {\red{that rely on a strongly gyrotropic \textit{permeability} tensor}} can be pushed up to a few hundreds of GHz using suitable materials \cite{shalaby2013magnetic}, but their magnetic response tails off at THz frequencies as their permeability converges to that of free-space. Moreover, at even higher frequencies, most magneto-optical materials {\red{(e.g., biased ferrimagnetic materials with a gyrotropic \textit{permittivity} tensor)}} exhibit very weak nonreciprocal properties \cite{zvezdin1997modern}, implying that large bias fields, or long propagation lengths (from tens to thousands of wavelengths), or narrow-band resonant structures, are required to achieve significant isolation \cite{du2018monolithic,zhang2019monolithic}.

Motivated by these outstanding challenges, in this paper we propose a new approach to realize broadband nonreciprocal effects in compact plasmonic structures. We revisit the dispersion characteristics of the permittivity model of a standard magnetized plasmonic material and unveil a previously overlooked regime of broadband and giant nonreciprocity below the cyclotron resonance frequency. This enables the realization of broadband isolators operating over more than a decade with low insertion loss and ultra-compact footprint, at the scale of the wavelength or smaller.

\section{II. Magnetized Plasma model}
We consider the simplest model of a nonreciprocal plasmonic medium, namely, a magnetized Drude plasma. For nonzero static magnetic bias, such a medium is anisotropic and gyrotropic, and its electromagnetic response is represented by a frequency-dispersive, asymmetric permittivity tensor $\epst$. For propagation along the direction of the bias (Faraday configuration), a diagonalization of $\epst$ shows that each circularly polarized (CP) eigenmode experiences a different effective permittivity $\eps_{\pm}$ (right-handed CP denoted by ``$-$'' and left-handed CP by ``$+$''). This creates a nonreciprocal phase shift between left-handed and right-handed CP waves if they are both allowed to propagate; therefore, a linearly polarized propagating wave experiences a polarization-plane rotation of $\Delta \Phi$ (Faraday rotation angle). To realize an isolator, it is then necessary to implement $\Delta \Phi = \pi/4$, such that a reflected wave undergoes a $\pi/2$ polarization rotation over a roundtrip, and insert a polarizer to filter out the rotated field at the input port.

\begin{figure} [h]
 \centering
 \includegraphics[ width=1\linewidth, keepaspectratio]{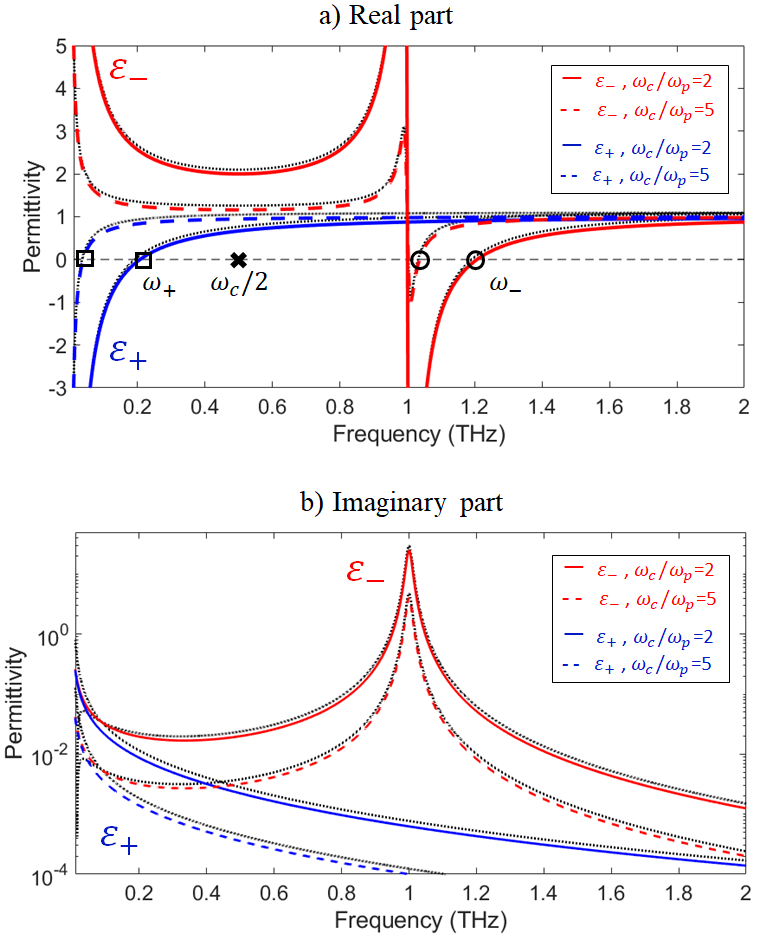}
 \caption{ Real  (a) and imaginary  (b) parts of the effective scalar permittivity for RCP ($\eps_{-}$) and LCP ($\eps_{+}$) modes, with $\wc/2\pi= 1$ THz, $\gamma=0.01\, \wc$, $\wc/\wp = 2\, (5)$ for solid (dashed) lines, assuming $\eps_{\infty}=1$.
 \red{The relative permittivity for the extraordinary mode below resonance, $\omega_c$, is always larger than unity, with minimum value at $\omega_c/2$, while for the ordinary mode the permittivity is smaller than unity.}  The dotted black lines represent the permittivity values derived using the modified Kramers-Kronig relations, shifted slightly upwards for clarity. The two regions of interest (in which both modes propagate with low loss) are: (i) the transparent region above $\w_{-}$ (small circle), and (ii) the transparent window, opened by the static bias, extending from $\w_{+}$ (small square) to $\wc/2$ (cross). The region below $\w_{+}$ is not of interest because the ordinary mode is an evanescent wave (negative permittivity); therefore, Faraday rotation is not possible.}
 \label{Fig1}
\end{figure}

The effective relative permittivity seen by the CP modes and the Faraday rotation angle (assuming free-space permeability) are given by (Ch. III.20 in Ref. \cite{sommerfeld1954lectures})
\begin{equation}
\eps_{\pm}(\w) =\epsinf \Big[ 1-\dfrac{\wp^2}{\w \,(\w \pm \wc +i \g)} \Big]
, 
\label{eq1}
\end{equation}
\begin{equation}
\Delta \Phi(\w) = \dfrac{L}{2 c_0} \, \w \,\,\Big(\Re \big[\sqrt{\eps_{-}(\w)}\big]-\Re\big[\sqrt{\eps_{+}(\w)}\,\big]\Big),
\label{eq2}
\end{equation}
where $\wp$ is the plasma frequency, $\wc = e B_{0} / m^{*}$ is the cyclotron frequency, with $B_{0}$ being the magnitude of the static bias field and $m^{*}$ the electron effective mass, $e$ is the electron charge, $\epsinf$ is the permittivity at high frequencies (bound-charge contribution), $\g$ is the loss coefficient, $L$ is the length of the medium, and $c_0$ is the free-space light speed. Throughout the paper, we assume a $e^{-i \w t}$ time-harmonic dependence for all fields.
As shown in Fig. \ref{Fig1}, $\eps_{-}$ (RCP mode) exhibits a resonance at $\wc$, corresponding to the frequency of the cyclotron motion of the plasma electrons under a static magnetic bias; this mode is denoted as the extraordinary mode. On the contrary, for the LCP mode, the material behaves as an ordinary Drude-like plasma, i.e., $\eps_{+}$ increases monotonically with frequency, albeit with a shifted plasma frequency. Frequencies at which $\eps_{\pm}$ crosses zero mark the limits of the propagation regions of interest, and are given by $\w_{\pm}= \sqrt{\wp^2+\wc^2/4} \mp \wc/2$.

In order to gain more physical insight into the dispersion characteristics of this permittivity model, it is crucial to appreciate how the principle of causality implies certain conditions and constraints, expressed through Kramers-Kroning (K-K) relations, which relate the real $\Re$ and imaginary $\Im$ parts of the permittivity function. Notably, using K-K relations, it is expected that $\partial \Re[\eps(\w)] / \partial \w >0 $ for passive materials in low-loss frequency windows (see section 84 in Ref. \cite{landau2013electrodynamics}, and the Appendix). However, as seen in Fig. \ref{Fig1}(a), while this condition applies to the LCP mode permittivity at all frequencies, it fails to describe the behavior of the RCP mode permittivity in the low-loss region below half the cyclotron  frequency. Therefore this may raise concerns about the applicability of K-K relations to gyrotropic plasmonic materials and, consequently, about the implications of causality for these media. 
In addition, we note that, for a nonzero cyclotron frequency (nonzero bias), the so-called ``reality condition'' (which implies a real time-domain response) is also violated in these materials: $\Re[{\eps_{\pm}(\w)}] \neq \Re[{\eps_{\pm}(-\w)}]$ and $\Im[{\eps_{\pm}(\w)}] \neq -\Im[{\eps_{\pm}(-\w)}]$.

While these apparent issues may seem surprising, we argue that they simply originate from the fact that $\eps_{\pm}$ are not elements of the permittivity tensor $\epst$, 
but eigenvalues that are combinations of diagonal and nondiagonal elements. Therefore the inverse Fourier transform of $\eps_{\pm}$ does not represent a ``true'' time-domain response function, and is not expected to be real \red{\cite{smith1976comments}}. Nevertheless, this fact does not prohibit the derivation of K-K relations, since the functions $\eps_{\pm}$ inherit the analyticity of the tensor elements (no poles in the upper half of the complex frequency plane). However, to obtain the correct K-K relations for $\eps_{\pm}$, it is crucial to properly take into account the contribution of the pole at $\w=0$ on the real frequency axis, which leads to an additional term for the K-K relation of the real part of the permittivity, compared to the standard K-K relation:

\begin{equation}
\Re[\eps_{\pm}(\w)] = 1+ \dfrac{1}{\pi} \dashint_{-\infty}^{\infty} d\W \dfrac{\Im[\eps_{\pm}(\W)]}{ (\W-\w)} \mp \dfrac{ \w_p^2 \w_c}{\w(\w_c^2+\gamma^2)}.
\label{eqKK}
\end{equation}

Further details on the modified K-K relations and their detailed derivation are available in  Appendix.

The results in Fig. \ref{Fig1} show that the permittivity functions (dotted black lines) derived from the modified K-K relations, Eqs. (\ref{eqKK}) and (\ref{modifiedKK}), exactly match the analytical permittivity expressions (\ref{eq1}) for both the real and imaginary parts, which confirms the causality of the standard permittivity model of a magnetized plasma. Most importantly, however, due to the aforementioned additional term in the K-K relation (\ref{eqKK}), the positive dispersion constraint, $\partial \Re[\eps(\w)] / \partial \w >0 $ in low-loss windows, is no longer guaranteed, as demonstrated by the extraordinary mode permittivity. In other words, anomalous (negative) dispersion is possible in this passive material, even in low-loss regions far from resonance. This makes the dispersion characteristics of a gyrotropic plasma fundamentally richer (nonmonotonic) than any isotropic materials. As discussed next, this nonmonotonic behavior can provide, under certain conditions, a natural form of dispersion compensation, and lays the foundations for achieving broadband giant nonreciprocity in magnetized plasmonic devices. 

\begin{figure} [h]
 \centering
 \includegraphics[ width=1\linewidth, keepaspectratio]{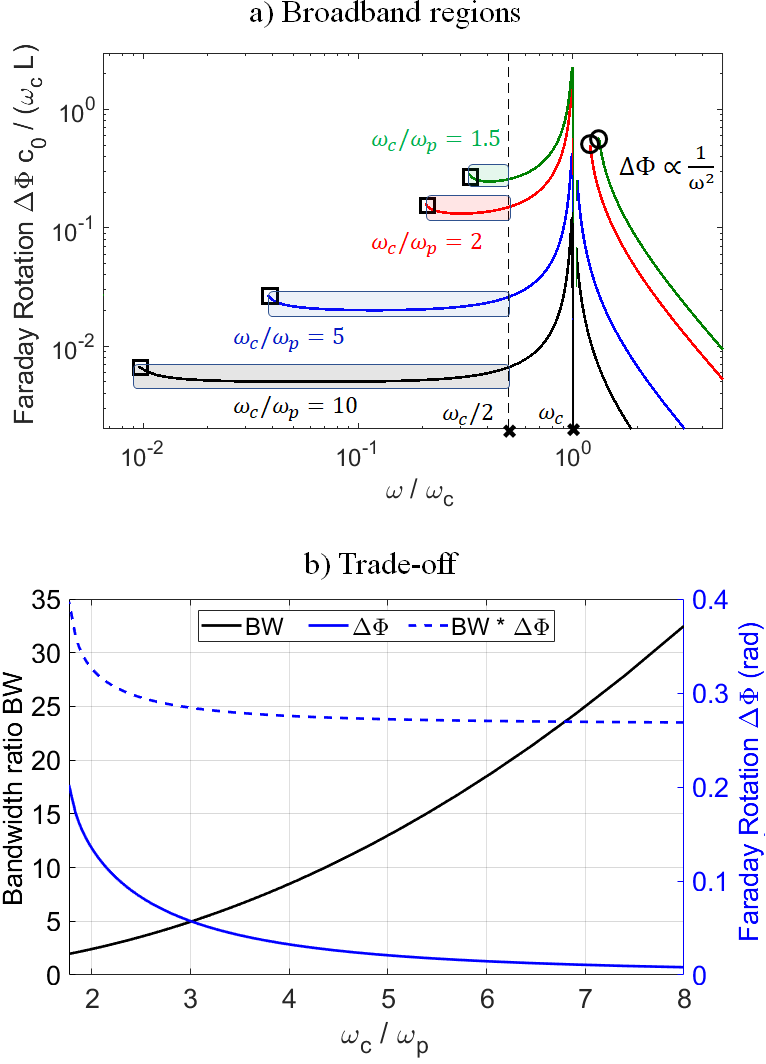}
 \caption{(a) Broadband regions of Faraday rotation (shaded areas) below the resonance frequency, with $\wc > 2 \wp / \sqrt{3}$, as discussed in the text. The different curves show the normalized Faraday rotation angle, $\Delta \Phi\, c_0 / (\red{\omega_c} L)$, for different values of $\omega_c/\omega_p$. We set  $\eps_{\infty}=1$ and $\gamma=0.01$. 
 (b) A fundamental trade-off exists between the width of the broadband region and the achievable Faraday rotation angle for different values of $\wc/\wp$. The width  of the broadband region BW is defined here as the ratio $(\wc/2)\, /\,(\w_{+})$. The dashed line represents the \textit{rotation-bandwidth} product, as defined in the text\red{, given  $a=0.04$ and $\eps_{\infty}=1$.} }
 \label{Fig2}
\end{figure}

\section{III. Broadband Faraday Rotation}
In order to achieve broadband Faraday rotation, with $\Delta \Phi(\w)$ approximately constant as a function of frequency, we need $\Big(\Re \big[\sqrt{\eps_{-}(\w)}\big]-\Re\big[\sqrt{\eps_{+}(\w)}\,\big]\Big) \propto	\w^{-1}$ over an extended frequency region, as deduced from Eq. (\ref{eq2}). Two candidate regions can be identified in Fig. \ref{Fig1}(a), where both permittivity functions are positive (propagating waves) and approach each other as frequency increases. The first region of interest is the transparent region above $\omega_{-}$, which is the most widely studied regime for Faraday rotation in plasmonic materials \cite{sommerfeld1954lectures,aplet1964faraday,weller2012optical,stadler2013integrated}. In this region, the permittivity function of both modes, $\eps_{\pm}$, monotonically increases with frequency and it approaches unity from below for $\w \rightarrow \infty$. Therefore, the Faraday rotation angle at high frequencies, where $\w \gg \wp,\wc, \g$, is approximately given by $\Delta \Phi \propto \w^{-2}$, as indicated in Fig. \ref{Fig2}(a) (this is the region conventionally associated with the definition of Verdet constant \cite{sommerfeld1954lectures}). Thus, in this regime, the material does not exhibit a broadband, dispersion-less, nonreciprocal behavior (conversely, due to their different microscopic material response, biased ferrites show nondispersive behavior in a similar ``far-from-resonance'' region, 
see, e.g., Eq. (7-18) in Ref. \cite{lax1962microwave}). 

The other region of interest, which has been overlooked so far in the literature, is below the cyclotron resonance, and is defined as the frequency range $\w_{+} \leq \w \leq \wc/2$, provided that the cyclotron frequency is sufficiently high, $\wc > 2 \wp / \sqrt{3}$, which assures the region existence as $\wc/2 > \w_{+}$. This corresponds to the anomalous-dispersion window in which $\partial \Re[\eps(\w)] / \partial \w <0 $ for the extraordinary mode. Note that, for any nonzero value of plasma frequency, $\wp \neq 0$, the extraordinary (ordinary) mode permittivity below resonance is always larger (smaller) than unity (or, more generally, the constant $\eps_{\infty}$) as seen in Fig. \ref{Fig1}(a). In the limit $\wc \gg \wp$, both permittivities, $\eps_{\pm}$, approach unity (or $\eps_{\infty}$) from opposite sides in this frequency window. 
Consequently, using a Taylor expansion, 
one arrives at $\Big(\Re\big[\sqrt{\eps_{-}(\w)}\big]-\Re\big[\sqrt{\eps_{+}(\w)}\,\big]\Big) \approx (\sqrt{\eps_{\infty}}\, \wp^2/\wc) \, {\w}^{-1}$. As discussed above, this is the desired frequency dependence to achieve broadband response: consistent with Eq. (\ref{eq2}), Faraday rotation is approximately dispersion-less in this region. The broadband behavior is demonstrated in Fig. \ref{Fig2}(a), where it is clear that the Faraday rotation angle gets less frequency dispersive as $\wc / \wp$ increases. We would like to stress, however, that broadband Faraday rotation with low dispersion can still be obtained even if $\wc$ and $\wp$ are comparable. For example, for an octave- (or a decade-) wide broadband region, it is required that $\wc / \wp > 1.8$ (or $4.4$), as shown in Fig. \ref{Fig2}(b). 

Attaining comparable cyclotron and plasma frequency such that $\wc > 2 \wp / \sqrt{3}$ may seem challenging (and it usually is) since it typically requires large magnetic bias. This issue may be circumvented using a metamaterial approach to lower $\wp$ \cite{pendry1996extremely}. Alternatively, $\wc > \wp $ is feasible in certain solid-state plasmas with small effective electron mass, as the ratio $\wc/\wp$ is proportional to 
$1/\sqrt{m^*}$. This has been experimentally demonstrated in, for example, $n$-doped Indium Antimonide (InSb) under moderate bias at THz frequencies \cite{wang2010interference,arikawa2012giant,wang2019photonic}. 

An interesting trade-off appears in Fig. \ref{Fig2} between the achievable Faraday rotation angle and the width of the broadband region. As $\wc / \wp$ increases, the difference between the permittivity values seen by RCP and LCP modes decreases, as shown in Fig. \ref{Fig1}(a) (compare the solid and dashed lines), which implies a smaller Faraday rotation angle. This can be quantified as follows: let the system length $L$ be defined in terms of the center wavelength of the considered band, that is, $L = a \lambda_s$, where $a$ is the normalized length, and $\ws = 2 \pi c_0 / \lambda_s $. Then, by letting $\ws=\wc/4$ (near the center of the band), the Faraday rotation angle can be approximately written as $ \Delta \Phi \approx 4.3 \pi \, a \sqrt{\eps_{\infty}}\,\wp^2 / \wc^2$. At the same time, increasing $\wc / \wp$ leads to a decrease of the lower-frequency limit, $\w_{+}$, of the broadband region, hence further broadening it. The bandwidth of the dispersion-less window in the $\wc \gg \wp$ limit can then be approximated as $BW \approx 0.5\,\wc^2/\wp^2$ \red{(bandwidth is defined here as the ratio of the high-frequency and the low-frequency limits (see Fig. \ref{Fig2}), as commonly done in microwave engineering to allow scalability across frequency bands of widely different absolute widths).} Thus, under the assumption of large $\wc / \wp$, we can define an approximate \emph{rotation-bandwidth product}, 
\begin{equation}
BW \times \Delta \Phi \approx 2.15\pi\, a \sqrt{\eps_{\infty}},
\label{eqRB}
\end{equation}
which quantifies the trade-off between maximum rotation and maximum bandwidth, with the bound uniquely determined by the length of the device and the high-frequency permittivity. This trade-off can be used as a metric for design optimization for broadband nonreciprocal devices operating in this new regime. We stress that, even with the limitations expressed by the rotation-bandwidth product, the Faraday rotation angles achievable in this below-resonance broadband region are significantly larger than their counterparts in the far-from-resonance high-frequency region, as can be seen in Fig. \ref{Fig2}(a). This enables the realization of not only broadband but also ultra-compact nonreciprocal components.

\section{IV. TEST CASE:  BROADBAND SUBWAVELENGTH THZ ISOLATOR}
As a rather striking example of broadband giant nonreciprocity in this new regime, we present a design for a deeply subwavelength ($a \ll 1$) broadband Faraday isolator operating at THz frequencies at room temperature (which is the first of its kind to our knowledge). 
As mentioned earlier, we choose $n$-type InSb due to its unparalleled potential for nonreciprocal devices at THz frequencies \cite{arikawa2012giant}, thanks to its low effective mass, high electron mobility, relatively low losses, and large $\eps_{\infty} \approx 15.68$ \cite{note}. 
The temperature dependence of $\wp(T)$, loss coefficient $\gamma(\w)$, and other parameters of InSb are adopted from Ref. \cite{mu2019tunable}. 

To design a broadband and efficient isolator, our goal is to achieve $ \Delta \Phi = \pi / 4$ within a $\pm 2.5\%$ margin (chosen arbitrarily) over the operational band, with power transmission above $85\%$ (insertion loss below $0.7$ dB), by optimizing the magnetic bias $B_{0}$ and the plasma frequency, while keeping the length $L$ of the device as small as possible. 
If the bias is fixed, the factor that primarily determines the bandwidth is the plasma frequency, which can be controlled through the temperature or doping.

\begin{figure} [h]
 \centering
 \includegraphics[ width=1\linewidth, keepaspectratio]{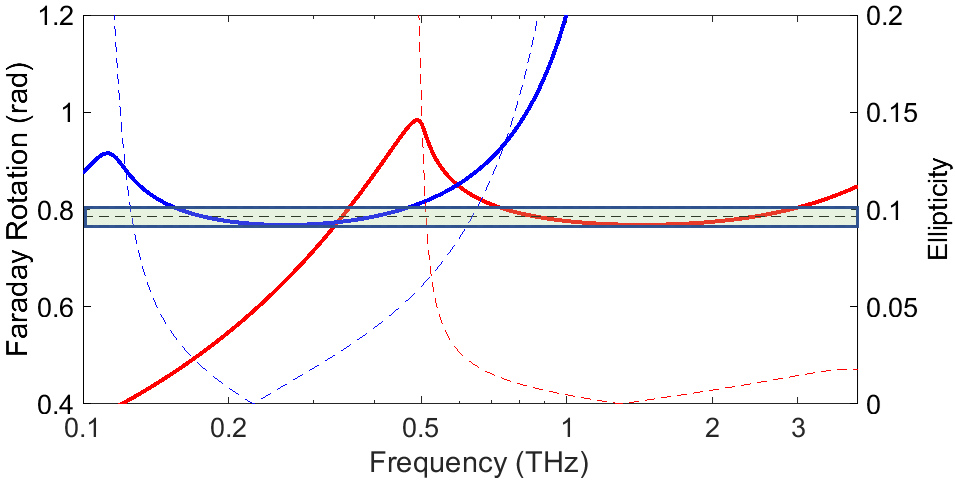}
 \caption{Two examples of \red{deeply} subwavelength broadband Faraday rotator at THz frequencies using $n$-type InSb: (red solid) moderately strong bias, $B_{0}=5\,$ T,  at room temperature, $290\, K$; and (blue solid) subtesla bias, $B_{0}=0.77\,$ T,  at $170\, K$.
 The rotation angle remains around $\pi/4$ (shaded area) across the operational band.  Dashed lines represents the ellipticity for each case (the smaller the better). The other parameters  are given in the text. }
 \label{Fig3}
\end{figure}

As an example, the red curve in Fig. \ref{Fig3} is for a  subwavelength slab of InSb with $L= \SI{33}{\micro\metre}$ ($a=0.192$), $B_{0}=5\,$ T, at room temperature $T= 290\, K$. 
 By using a suitable polarizer, this subwavelength structure can then work as a broadband and efficient isolator that covers the $0.7-3$ THz band. To cover the lower frequencies in the high-GHz regime, a lower plasma frequency is required, which can be obtained by reducing the doping or the temperature. The blue curve in Fig. \ref{Fig3} is for another subwavelength Faraday rotator with $L= \SI{0.138}{\milli\metre} $ ($a=0.15$), and a much lower bias, $B_{0}=0.77$ T, at $T= 170\, K$, which covers the band $0.15$-$0.45$ THz. Both cases in Fig. \ref{Fig3} exhibit the required Faraday rotation within the specified margin of $\pm 2.5\%$ and with transmitted power above $85\%$ (due to the presence of unavoidable losses in the plasmonic material), assuming matched isolator ports using suitable anti-reflection coatings. 
 
 For completeness, in Fig. \ref{Fig3}, we also characterize the isolator dichroism (polarization-dependent absorption) based on the ellipticity parameter, which is typically defined as the difference between the amplitudes of the two propagating CP waves divided by their sum. Ellipticity needs to be minimized to ensure that the linearly-polarized state of the wave is preserved as it propagates through the Faraday isolator, so that a reflected wave is efficiently absorbed by the polarizer. As shown in Fig. \ref{Fig3}, the ellipticity is negligible for both cases across the indicated operational bands.


\section{V. Conclusion}

By revisiting the dispersion properties of gyrotropic plasmonic materials and having clarified the implications of causality in this case, we have unveiled a previously overlooked regime of wave propagation in biased plasmas, characterized by anomalous dispersion and low losses. This leads to a broadband window of dispersion-less Faraday rotation, which could extend over more than a decade-wide frequency band. Contrary to the conventional regime of operation of magnetized plasmas and ferrites, this broadband window is closer to resonance and, therefore, is characterized by stronger nonreciprocity, hence enabling the realization of deeply subwavelength nonreciprocal devices, with an orders-of-magnitude reduction in length for the same level of bias intensity. 

The results presented in this paper may pave the way toward superior nonreciprocal components that can be optimized for both broadband operation and compactness. We expect our findings to be particularly appealing for various emerging applications within the wide ``THz gap,'' where efficient nonreciprocal devices based on biased ferrites or spatio-temporally modulated structures are difficult to realize.

\appendix
\section{APPENDIX: KRAMERS-KRONIG (K-K) RELATIONS FOR GYROTROPIC PLASMONIC MEDIA}

Certain physical constraints exist that restrict the response function of any physical system, regardless of its specific details. For instance, causality ensures the analyticity of the permittivity function $\eps(\w)$ in the upper half of the complex frequency plane. Using the residue theorem, this property leads to the following relation:
\begin{equation}
 \oint_{\Gamma} d\W \, \dfrac{\eps(\W)-1}{\W-\w} = 0,
 \label{eq:cauchy}
\end{equation}
where $\Gamma$ is an arbitrarily closed contour in the upper half plane. For the plasma model used in this paper, $\eps(\w)$ has a pole on the real axis at $\w=0$. Thus, let $\eps(\w)-1 = \chi(\w)= \c0 / \w$, where $\c0$ is analytic in the upper half plane and on the entire real axis. \red{An alternative, equivalent method to derive K-K relations for this case is to evaluate the contour integral for the function $\omega  \chi(\w) $, to avoid the pole at $\omega=0$ \cite{bulik1995polarization,potekhin2004electromagnetic}}.

\begin{figure} [h]
 \centering
 \includegraphics[ width=0.9\linewidth, keepaspectratio]{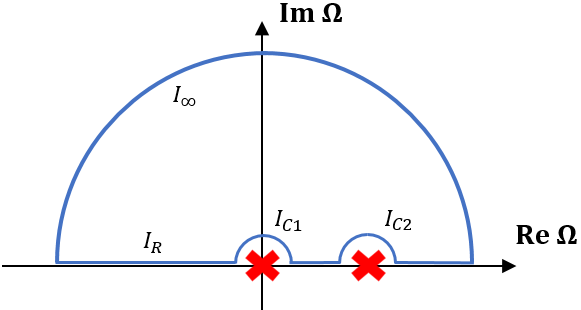}
 \caption{The integral path $\Gamma$ in the complex frequency plane used to derive the modified Kramers-Kroning relations for the effective permittivity of  gyrotropic plasmas.}
 \label{FigApp}
\end{figure}

Now let the closed integration path $\Gamma$ be an infinite semicircle in the upper half plane, excluding the poles at $\Omega =0$ and $\Omega=\w$, i.e., $\Gamma = I_{\infty} + I_R + I_{C1}+I_{C2} $, as shown in Fig. \ref{FigApp}. For such a causal system, and given that the energy of any physical interaction is necessarily finite, it can always be ensured through Titchmarsh theorem that $ \chi_0(\omega) \to 0$ as $\Re[\omega] \to \pm \infty$ \cite{fearn2006dispersion} and the integral over the semi-infinite circle $I_{\infty}$ evaluates to zero.

Then, by evaluating the two infinitesimal semicircles $I_{C1}$ and $I_{C2}$ around the poles, Eq. (\ref{eq:cauchy}) becomes
\begin{equation}
 \dashint_{-\infty}^{\infty} d\W \dfrac{\chi_0(\W)}{\W \, (\W-\w)} + i \pi \dfrac{\chi_0(0)}{\w} - i \pi \dfrac{\chi_0(\w)}{\w } = 0,
\end{equation}
where the integral $\dashint$ is the Cauchy Principle-Value integral defined over $\mathbb{R}-\{0,\w\}$, represented by $I_R$. Crucially, $\w \neq 0$ is required, otherwise, the integrand would have a second-order pole at $\W = 0$, and such a PV integral diverges for second-order poles. It is also important to stress that $\dashint d\Omega/\Omega$ must be approached symmetrically to be correctly defined (and so that it converges) \cite{arfken1999mathematical}.

By rearranging, we arrive at the modified K-K relations for magnetized plasmonic materials valid for both $\eps_{\pm}$ (for $\w \neq 0$),
\begin{subequations}
\label{modifiedKK} 
\begin{eqnarray}
\Re[\eps(\w)] = 1+ \dfrac{1}{\pi} \dashint_{-\infty}^{\infty} d\W \dfrac{\Im[\eps(\W)]}{ (\W-\w)} + \dfrac{\Re[\chi_0(0)]}{\w},\label{eq5}
\\[8pt]
\Im[\eps(\w)] = \dfrac{-1}{\pi} \dashint_{-\infty}^{\infty} d\W \dfrac{\Re[\eps(\W)-1]}{ (\W-\w)} + \dfrac{\Im[\chi_0(0)]}{\w}.\label{equationb}
\end{eqnarray}
\end{subequations}

For the case of a non-magnetized plasma described by a scalar lossy Drude model, $\Re[\chi_0(0)] = 0 $, and since $\Im[\eps(\W)]$ is positive at all frequencies in passive media, it is straightforward to show that the frequency derivative of the real part of the permittivity (\ref{eq5}) is always positive in low-loss frequency windows \cite{landau2013electrodynamics}, corresponding to normal dispersion. This is not valid in the case of gyrotropic plasmas because $\Re[\chi_0(0)] = \mp \w_p^2 \w_c / (\w_c^2+\gamma^2) \neq 0$; therefore, it is easy to see that, for the extraordinary mode, the normal-dispersion constraint $\partial \Re[\eps(\w)] / \partial \w >0 $ can no longer be assumed to be valid in low-loss windows. In particular, in any low-loss region, we get
\begin{equation}
 \dfrac{ \partial \Re[\eps_{\pm}(\w)]}{ \partial \w} = \dfrac{1}{\pi} \dashint_{-\infty}^{\infty} d\W \dfrac{\Im[\eps_{\pm}(\W)]}{ {(\W-\w)}^2} - \dfrac{\Re[\chi_0(0)]}{\w^2},
\end{equation}
from which it is clear that the negative term is dominant  at low frequencies for the extraordinary mode (corresponding to the anomalous dispersion region discussed in the main text).

\begin{thebibliography}{39}
\bibitem [{not()}]{note}%
  \BibitemOpen
  \href@noop {} {}\bibinfo {note} {The InSb permittivity model at THz
  frequencies includes an additional resonance term due to phonon contribution
  at $5.9$ THz. We limit our investigation to frequencies below this resonance
  to avoid an additional source of losses.}\BibitemShut {Stop}%
\end{thebibliography}

\providecommand{\noopsort}[1]{}\providecommand{\singleletter}[1]{#1}%

\end{document}